\documentclass[12pt,a4paper,english]{article}
\usepackage[T1]{fontenc}
\usepackage[latin2]{inputenc}
\usepackage[english]{babel}
\usepackage{amsmath}
\usepackage{amsfonts}
\usepackage{indentfirst}
\usepackage[dvips]{graphicx}

\newcommand{\bea}{\begin{eqnarray}}
\newcommand{\eea}{\end{eqnarray}}
\newcommand{\be}{\begin{equation}}\begin{tiny}\end{tiny}
\newcommand{\ee}{\end{equation}}

\def\Tr{{\rm Tr}}

\def\G{\Gamma}

\newcommand{\Li}{{\rm Li}}

\begin{document}

\sloppy


\begin{flushright}
\begin{tabular}{l}
CALT 68-2767 \\

\\ [.3in]
\end{tabular}
\end{flushright}

\begin{center}
\Large{ \bf Matrix models for $\beta$-ensembles \\
from \\
Nekrasov partition functions}
\end{center}

\begin{center}

\bigskip

\bigskip 

Piotr Su{\l}kowski$^{\,}$\footnote{On leave from University of Amsterdam and So{\l}tan Institute for Nuclear Studies, Poland.}

\bigskip

\bigskip

\emph{California Institute of Technology, Pasadena, CA 91125, USA} \\ [1mm]

\bigskip


\bigskip

\smallskip
 \vskip .4in \centerline{\bf Abstract}
\smallskip

\end{center}

We relate Nekrasov partition functions, with arbitrary values of $\epsilon_1,\epsilon_2$ parameters, to matrix models for $\beta$-ensembles. We find matrix models encoding the instanton part of Nekrasov partition functions, whose measure, to the leading order in $\epsilon_2$ expansion, is given by the Vandermonde determinant to the power $\beta=-\epsilon_1/\epsilon_2$. An additional, trigonometric deformation of the measure arises in five-dimensional theories. Matrix model potentials, to the leading order in $\epsilon_2$ expansion, are the same as in the $\beta=1$ case considered in 0810.4944 [hep-th]. We point out that potentials for massive hypermultiplets include multi-log, Penner-like terms. Inclusion of Chern-Simons terms in five-dimensional theories leads to multi-matrix models. The role of these matrix models in the context of the AGT conjecture is discussed.


\newpage 

\tableofcontents

\newpage 


\section{Introduction and summary}

Quantum gauge theories with $\mathcal{N}=2$ supersymmetry appear to possess a rare feature of getting more and more fascinating as the time flows by. Since the seminal solution of Seiberg and Witten \cite{SW-1,SW-2}, it is known that their solution is encoded in Seiberg-Witten curves. The very existence of these curves inspired several important developments and led to remarkable relations to other branches of physics and mathematics. 

The first such branch is a theory of random partitions. In \cite{Nek,Nek-Ok,NY-I,NY-II,ABCD} it was shown that Seiberg-Witten curves arise as limiting shapes of large ensembles of random two-dimensional partitions. These partitions label instanton configurations, which are the only contributions to the gauge theory partition function computed by localization. To apply localization techniques one has in fact to consider a more general setup, which, among several other subtleties, involves a non-trivial spacetime metric called the $\Omega$-background. This background is characterized by two parameters $\epsilon_1$ and $\epsilon_2$. Ultimately one can explicitly write down partition functions of gauge theories $Z$, which now bear the name of Nekrasov, and in particular they depend on $\epsilon_1$ and $\epsilon_2$. The information about the Seiberg-Witten curve is encoded in the prepotential $F_0$, which is the leading term of the free energy $F_0=-\lim_{\epsilon_1,\epsilon_2\to 0} \epsilon_1\epsilon_2 \log Z$. In the special case $\epsilon_1=-\epsilon_2=\hbar$ the free energy has the expansion 
\be
F= \log Z = \sum_{g=0}^{\infty} F_g \hbar^{2g-2},     \label{FlogZ}
\ee
and higher genus terms $F_g$ are also associated to the ordinary Seiberg-Witten curve. For arbitrary $\epsilon_1,\epsilon_2$ 
this expansion takes more general form and it is natural to expect that it should be associated to some deformation of the Seiberg-Witten curve. As we explain further there are strong indications that the non-commutative deformation arises in such a general case. 

Another class of systems, whose solutions are encoded in complex curves, are matrix models, defined by matrix integrals  
\be
Z =  \int \mathcal{D}M e^{-\frac{1}{\hbar} \Tr V(M)}.     \label{Zmatrix}
\ee
The integration measure can typically be diagonalized and expressed in terms of eigenvalues $x_i$ of matrices $M$. In case of hermitian matrices, such diagonalization results in the Vandermonde determinant $\mathcal{D}M=\prod_i dx_i \prod_{i<j}(x_i-x_j)^{2}$. The solution of such matrix integrals can also be expressed in the form (\ref{FlogZ}), which is encoded in the underlying spectral curve. One is therefore tempted to devise such (hermitian) matrix models, whose spectral curves would be identified with Seiberg-Witten curves, and moreover the full solutions (\ref{FlogZ}) on both sides would match. Such a program was initiated by Dijkgraaf and Vafa in \cite{DV-0206,pert-window}, where they related both systems by a chain of dualities which involved topological string theory. The connection to topological strings is not a surprise: their amplitudes also take form (\ref{FlogZ}), and it is known that $\mathcal{N}=2$ theories can be engineered by considering topological strings on non-compact Calabi-Yau manifolds \cite{geom-eng}, whose geometry is encoded in yet another complex (B-model) curve \cite{adkmv}. This relation is exact, i.e. topological string partition functions agree with Nekrasov partition functions to all orders in genus expansion \cite{SUNamplitudes,HIV}. This agreement was also extended, in terms of refined topological vertices \cite{refined-vertex,refined-Kanno}, to arbitrary values of $\epsilon_1,\epsilon_2$ in \cite{Taki,AwataKanno-refined}.

However, the relation of Seiberg-Witten (and therefore topological strings) to matrix models is in fact more subtle: while it was shown that the expansions (\ref{FlogZ}) agree for the lowest order terms $F_0,F_1$ even for simple polynomial potentials \cite{gaugedmm,kmt}, the agreement of the full series (\ref{FlogZ}) is much harder to achieve and depends on details of matrix potentials $V(M)$ \cite{kmr}. Nonetheless, recently matrix models have been found \cite{SW-matrix,matrix2star}, which by the very construction agree with the Nekrasov partition functions, and therefore automatically lead to the identification of full expansions (\ref{FlogZ}) on both sides.

Recently the above web of dualities has been extended again, in a novel and surprising direction: it was observed that Nekrasov partition functions agree with correlators in Liouville theory \cite{AGT}. In view of the connections described above, one might therefore expect that there are connections between Liouville theory and matrix models. Such connections have been proposed recently by Dijkgraaf and Vafa \cite{DV2009}, in a way which extends their earlier ideas mentioned above. We will discuss these relations more in what follows.

\bigskip

In the above web of dualities matrix model possess one peculiar feature. While the amplitudes in gauge theories, topological strings,\footnote{Topological strings, as such, are defined for a single parameters $g_s=\epsilon_1=-\epsilon_2$. However their amplitudes can be refined in a natural, combinatorial way to two parameters using the construction of refined vertices \cite{refined-vertex,refined-Kanno}. It is argued in \cite{AwataKanno-refined}, that for topological strings on toric Calabi-Yau manifolds which engineer gauge theories, the amplitudes computed using these two vertices give the same results.} and Liouville theory depend in a natural way on the two parameters $\epsilon_1, \epsilon_2$, it is not a priori clear how (or whether at all) the dependence of the matrix integral (\ref{Zmatrix}) on $\hbar$ can be refined. Nonetheless, matrix models are essential ingredients of all those dualities, and therefore they would appear incomplete without such a refinement. 

In this paper we show that such a refinement of matrix models indeed exists, and it takes form of the so-called $\beta$-deformation. By explicit computation, we transform a large class of Nekrasov partition functions into the form of $\beta$-deformed matrix models, i.e. matrix models of the form (\ref{Zmatrix}), however with modified determinant \cite{Mehta,beta-matrix-1,beta-matrix-2}. In the four dimensional case this determinant is given by
\be
\qquad \qquad \mathcal{D}M=\prod_i dx_i \prod_{i<j}(x_i-x_j)^{2\beta},     \label{betaVand}
\ee
and the exponent $\beta$ captures the dependence on parameters of the $\Omega$-background 
\be
\beta = - \frac{\epsilon_1}{\epsilon_2}.
\ee
Clearly the special case $\epsilon_1=-\epsilon_2=\hbar$, or $\beta=1$, corresponds to the ordinary Vandermonde determinant. We derive $\beta$-deformed matrix models for four- and five-dimensional $U(n)$ theories with massive hypermultiplets, as well as five-dimensional Chern-Simons terms. In the five-dimensional case the determinant (\ref{betaVand}) is in addition deformed in the trigonometric way. The generalization to six dimensions is straightforward, and results in the elliptic deformation, similar as in \cite{HIV,matrix2star}. The methods which we use are essentially extension of those introduced in \cite{SW-matrix,matrix2star,eynard-planch}; for related developments see also \cite{matrixHurwitz,MatrixW,equivalenceHurwitz,matrixPlane}.

It was already argued in \cite{DV2009} that refinement of matrix models should take such a form, based on properties of the $\Omega$ background, as well as the example of the $\beta$-deformed Gaussian matrix model. This Gaussian model is a special case of the Selberg integral, and its exact value is known. However this example seems to relate only to four-dimensional theories, and it is not clear how to extend it beyond the quadratic potential. 

More precisely, the $\beta$-deformed matrix models in the form given above, arise in our computation in the leading order in the expansion in one of the parameters, say $\epsilon_2$. The $\beta$-deformation of the measure is extracted from the asymptotics of its yet more general form. In addition there is the whole subleading series in $\epsilon_2$, both in the measure, as well as in the matrix model potential. This is analogous to the results in \cite{SW-matrix,matrix2star,eynard-planch,matrixHurwitz} for $\beta=1$, where matrix potentials (but not the measure) included additional subleading series in $\hbar$. An additional subtlety in the present $\beta\neq 1$ case is that these subleading terms cannot be immediately symmetrized (in a sense which will be explained in what follows), which makes their matrix model interpretation less clear. Nonetheless, we postulate that these subleading terms are essentially inessential, and the knowledge of just the leading terms should be sufficient to analyze the entire theory. This is so in matrix models corresponding to $\beta=1$ analyzed in \cite{SW-matrix,matrix2star,eynard-planch}. A detailed explanation of this phenomenon is given in \cite{matrixHurwitz}, where analogous matrix models for Hurwitz numbers are derived. In these cases the leading piece of the matrix model is sufficient to solve the whole matrix model, because it encodes the spectral curve, and the knowledge of the spectral curve is sufficient to define recursively symplectic invariants of Eynard and Orantin \cite{eyn-or}. In case of curves arising in the B-model topological strings, these symplectic invariants agree with Gromov-Witten invariants \cite{remodel,HKII}. In fact, the results of \cite{eyn-or} have been recently extended to the matrix models for $\beta$-ensembles \cite{beta-matrix-1,beta-matrix-2}. We therefore suppose that similar scenario should hold in the present case.

\bigskip

The organization of this paper is as follows. In the reminder of this section we summarize in detail our results, and discuss their relation to the AGT conjecture. In section \ref{sec-SW-Nekrasov} we present Nekrasov partition functions in the form appropriate for our purposes. In section \ref{sec-matrixmodel} we present a general scheme of deriving matrix models associated to Nekrasov partition functions. In section \ref{sec-4d} we provide such a derivation for four-dimensional theories. In section \ref{sec-5d} we provide a derivation for five-dimensional theories. In section \ref{sec-discussion} we discuss directions for further studies. In appendices we present yet more details on Nekrasov partition functions, explain manipulations with infinite sums, present various asymptotics necessary to extract $\beta$-deformed measures, and provide a complete example of the subleading terms in the four-dimensional matrix model potential.


\subsection{Summary of the results}

In this paper we find $\beta$-deformed matrix models, which encode Nekrasov partition functions for theories $U(n)$ gauge groups, with various matter contents and in various dimensions. We find it convenient to rescale the constant in front of the potential in (\ref{Zmatrix}), and write the $\beta$-deformed matrix models in the form 
\be
Z =  \int_{Mat_{nN}} \mathcal{D}M e^{-\frac{1}{\epsilon_2} \Tr V(M)},    \label{Zmatrix2}
\ee
where $M\in Mat_{nN}$ denotes $nN\times nN$ matrices from $\beta$-ensembles.

For four-dimensional theories, to the leading order in $\epsilon_2$, we find that the measure $\mathcal{D}M$ involves the $\beta$-deformed Vandermonde determinant
\be
\mathcal{D}M=\prod_i dx_i \prod_{i<j}(x_i-x_j)^{2\beta},
\ee
while four-dimensional potentials take form
$$
V^{4d}(x) = tx + V^{4d}_{vec}(x) + V^{4d}_{(anti)fund}(x).
$$
The linear term $tx$ encodes a dependence on the instanton counting parameter, while the contributions from vector and hypermultiplets in the fundamental representations, to the leading order in $\epsilon_2$, are given by
\bea
V^{4d}_{vec}(x) & = & 2\sum_{l=1}^n \Big( (x-a_l)\log(x-a_l) - (x-a_l) \Big), \nonumber \\
V^{4d}_{fund}(x) & = & \sum_{{\bf f} =1}^{N_{\bf f}} \Big( -(x - m_{\bf f})\log(x - m_{\bf f}) + (x - m_{\bf f}) \Big). \nonumber 
\eea
The expression for $V^{4d}_{antifund}(x)$ is analogous to $V^{4d}_{fund}(x)$, but with masses $m_{\bf f}$ replaced by $\epsilon_1+\epsilon_2-m_{\bf f}$. We note in particular that potentials for matter in (anti)fundamental representation include Penner-like factors of the form 
$$
\sum_{\bf f} m_{\bf f}\log(x-m_{\bf f}).
$$ 
Analogous factors were obtained from completely different perspective by Dijkgraaf and Vafa in \cite{DV2009}. While the arguments of logarithms in \cite{DV2009} were chosen, at least to some extent, at will due to conformal invariance, in our case they are fixed as $(x-m_{\bf f})$.

We also derive matrix models for 5-dimensional theories compactified on a circle of radius $R_5$. We usually suppress the dependence on this radius, and it can be reintroduced by rescaling $x$ and other parameters (such as $a_l,m_{\bf f}$) by $R_5$. The matrix models for five-dimensional theories also take form (\ref{Zmatrix2}), with the measure which involves, to the leading order in $\epsilon_2$, the Vandermonde determinant with both trigonometric and $\beta$-deformation
\be
\mathcal{D}M = \prod_i d x_i  \prod_{i<j} \Big(2 \sinh \frac{x_i - x_j}{2}\Big)^{2\beta}.
\ee
The five-dimensional potentials read
$$
V^{5d}(x) = tx + V^{5d}_{vec}(x) + V^{5d}_{(anti)fund}(x). 
$$
The linear factor $tx$ encodes in particular the dependence on the instanton counting parameter, while contributions from vector and hypermultiplets, to the leading order in $\epsilon_2$, are given respectively by
\bea
V^{5d}_{vec}(x) & = & \frac{n}{2} x^2 + 2\sum_{l=1}^n \Li_2(e^{-x+a_l}), \nonumber \\
V^{5d}_{fund}(x) & = & \sum_{\bf f} \Li_2\big(e^{x-m_{\bf f}}\big).
\eea
The contribution for antifundamental matter $V^{5d}_{antifund}(x)$ is related to $V^{5d}_{fund}(x)$ as in the four-dimensional case. 

In five dimensions one can also introduce Chern-Simons terms, which are parametrized by a single integer $m_{CS}$. In this case we obtain multi-matrix models with additional linear terms
$$
V^{5dCS} = m_{CS}\sum_{l=1}^n \Big(a_l \sum_{i=1}^N x^{(l)}_i \Big),
$$
where $x^{(l)}_i$ represent eigenvalues of the $l$'th set of matrices.

We stress that all these potentials are the same as in the $\beta=1$ case discussed in \cite{SW-matrix}. In the present case the only modification arises as the $\beta$-deformation of the Vandermonde determinant.


\subsection{Relation to the AGT conjecture}

As we already mentioned, in \cite{AGT} remarkable connections between four-dimensional Seiberg-Witten theories and two-dimensional conformal field theories have been conjectured. This observation was again motivated by properties of Seiberg-Witten curves: in \cite{GaiottoN2} they were related to other curves, whose gluing allows to construct more complicated gauge theories from simpler ones, in a way analogous to the construction of conformal field theories on arbitrary Riemann surfaces. Then direct comparison of certain quantities in several theories from both these classes led to general conjectures. These quantities involve one-loop and instanton parts of Nekrasov partition functions $Z^{1-loop},Z^{inst}$ on the Seiberg-Witten side, and Liouville three-point functions $C^{DOZZ}_{123}$ given by the DOZZ formula and conformal blocks $\mathcal{F}_{12}^{34}$ on the CFT side.

In particular, under this correspondence, the central charge of the Liouville theory 
$$
c=1+6\big(b+\frac{1}{b}\big)^2, \qquad \qquad b^2 = \frac{\epsilon_1}{\epsilon_2} = -\beta,
$$ 
is expressed in terms of the ratio of $\epsilon_1$ and $\epsilon_2$. In the special case $\epsilon_1=-\epsilon_2=\hbar$ we obtain $c=1$ and the Liouville theory reduces to the free fermion theory. Further aspects and checks of this correspondence were discussed, among the others, in \cite{
WyllardAGT,MirMor-Power,MirMor-U3,MarMirMor-Nf2Nc,MirMor-c,AGGTV-surface,DGOT-loop,qVirAGT,Quiver2d4d,PennerEguchi,SchiappaWyllard,MorozovBS,TaiGenusOne}. 

\begin{figure}[htb]
\begin{center}
\includegraphics[width=0.85\textwidth]{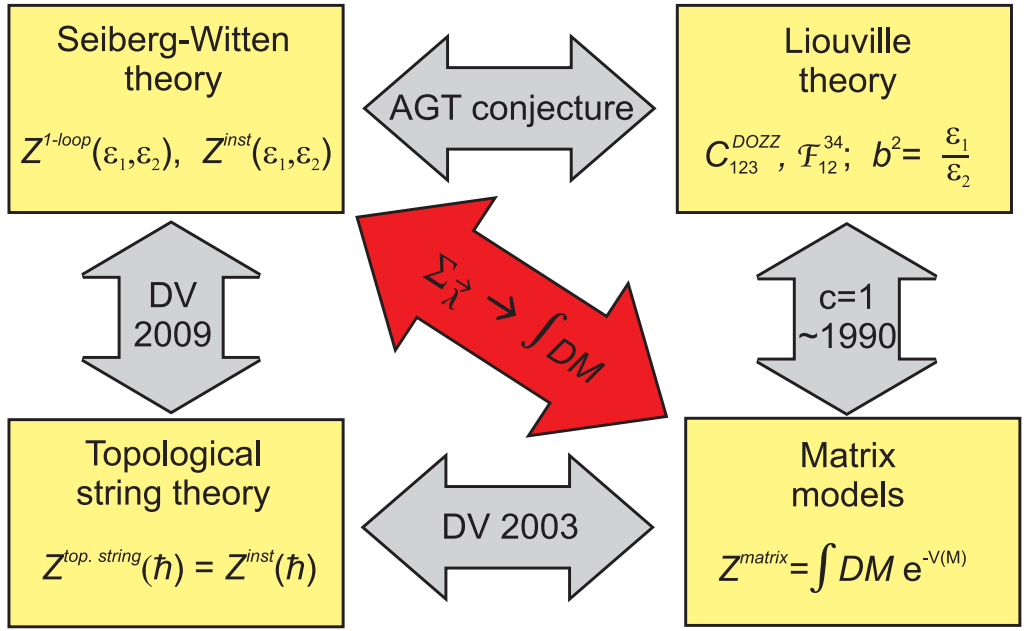} 
\begin{quote}
\caption{\emph{Dualities relevant to the AGT conjecture. The top grey arrow represents the original AGT conjecture \cite{AGT}. The other grey arrows represent a chain of dualities proposed by Dijkgraaf and Vafa \cite{DV2009}. The red arrow represents our present explicit reformulation.}} \label{fig-AGT}
\end{quote}
\end{center}
\end{figure}

Dijkgraaf and Vafa proposed a derivation of this conjecture inspired by topological string dualities and relation to matrix models \cite{DV2009}. A sequence of these dualities is shown in figure \ref{fig-AGT}. The top arrow represents the original AGT conjecture \cite{AGT}. The bottom arrow refers to general relations between topological strings and matrix models discovered by Dijkgraaf and Vafa in \cite{DV-0206,pert-window}. The crucial ingredient of their recent derivation \cite{DV2009} involves engineering Calabi-Yau manifolds relevant for theories considered in \cite{AGT}, and is represented by the vertical arrow on the left. The chain of dualities is completed by the arrow on the right, representing the relation between matrix models and conformal field theories \cite{MMM-CFT,Kostov}.
In fact, to make this chain of dualities rigorous, several issues are still to be completed. For example, the relation between matrix models and conformal field theories is known to hold rigorously only in $c=1$, or equivalently $\beta=1$ case. Also the relations to topological strings discussed in \cite{DV2009} were provided for $\beta=1$ case, and their extension beyond this case is more subtle. Let us also note, that various aspects of the matrix models proposed in \cite{DV2009} were further analyzed in \cite{Quiver2d4d,PennerEguchi,SchiappaWyllard,MorozovBS,TaiGenusOne}.

We wish to point out, that our present results can be regarded as an explicit realization of a part of the program proposed in \cite{DV2009}. Our results are represented by the red arrow in figure \ref{fig-AGT}: we directly relate Nekrasov partition functions to $\beta$-deformed matrix models. As argued earlier, we suppose that the leading form of these matrix models should be sufficient to solve entire theory, due to powerful recursive relations found in \cite{beta-matrix-1,beta-matrix-2,eyn-or}. (In fact, due to equivalence of five-dimensional gauge theories and topological strings \cite{SUNamplitudes,HIV,Taki,AwataKanno-refined}, our results automatically realize the equivalence represented by the bottom arrow, which therefore could also be drawn in red.)

Furthermore, the relation to Liouville theory might be completed due to the equivalence of non-commutative (or quantum) curves arising on both sides.
It was proposed in \cite{AGT} how to associate a quantum curve to the Liouville theory. This curve is given by the operator equation
$$
x^2 = T(z),
$$ 
where $T(z)$ is the Liouville energy-momentum tensor, and it is supposed to encode the data of the Liouville theory. On the other hand, the non-commutative spectral curves have been recently associated to $\beta$-deformed matrix models in \cite{beta-matrix-1,beta-matrix-2}, and it was shown how to generalize to this case the recursion relations familiar from the $\beta=1$ case \cite{eyn-or}. 
Even though matrix models considered in \cite{beta-matrix-1,beta-matrix-2} have only polynomial potentials, one might hope to extend those results to the potentials which we obtain here. In such case, analogous non-commutative curves might be associated to our matrix models. The adventage of such an approach would be twofold. Firstly, the corresponding non-commutative symplectic invariants should hopefully agree with refined Seiberg-Witten and (for five-dimensional theories) refined Gromov-Witten invariants. This would provide more rigorous proof of the AGT conjecture. Secondly, this would provide a concrete proposal for the non-commutative generalization of the Seiberg-Witten curve -- i.e. if such a generalization exists, it is natural to expect that it should agree with the (non-commutative) matrix model curve, similarly as in $\beta=1$ case \cite{remodel}. It would be very interesting to understand such a generalization of the Seiberg-Witten curve purly from the gauge theory perspective, as well as from its embedding in string theory. In this regard we note that yet another form of quantum curves have been considered in the context of  Seiberg-Witten theories in \cite{dhsv,dhs}, which have a string theory interpretation in terms of intersecting D4- and D6-branes. Supposedly those curves should also be related to the non-commutative curves mentioned above, and the B-field present in the intersecting brane system would provide physical reason for the non-commutativity. The relations between matrix models and non-commutative curves are discussed also in \cite{Quiver2d4d}. Connections between all these issues are currently under investigation.


\section{$\mathcal{N}=2$ theories and Nekrasov partition functions}          \label{sec-SW-Nekrasov}

Partition functions for $\mathcal{N}=2$ theories were derived by Nekrasov using localization techniques in \cite{Nek} and analyzed 
further in great detail, among the others, in \cite{Nek-Ok,NY-I,NY-II,AwataKanno-refined,Taki,AGT}. They consists of the one-loop and instanton part. In this paper we consider only instanton parts, and refer to them simply as \emph{(Nekrasov) partition functions}.

Partition functions for $U(n)$ theories can be viewed as a generalization of the Plancherel measure to the case of $n$ sets of partitions \cite{Nek-Ok,SW-matrix}. They are given by a sum over a set of $n$ partitions $\lambda^{(l)}_i$, with 
$l=1,\ldots,n$ labeling various partitions and index $i > 0$ denoting the $i$'th row of a given partition. 
These partition functions depend the scalar vevs $a_l$ via $a_{lk}=a_l-a_k$, the dynamically generated scale or other counting parameter $\Lambda$, and the parameters of the $\Omega$-background $\epsilon_1,\epsilon_2$. We also introduce
\be
\beta = - \frac{\epsilon_1}{\epsilon_2}
\ee
which will play an important role from the matrix model perspective. Of course a dependence on masses $m_{\bf f}$ arises in theories with $N_{\bf f}$ massive hypermultiplets.
We introduce also
$$
b_l = \frac{a_l}{\epsilon_2},\qquad \qquad M_{\bf f} = \frac{m_{\bf f}}{ \epsilon_2}.
$$ 
In five-dimensional theories there is also a dependence on the radius of the fifth dimension $R_5$, which we usually suppress. It can be reintroduced by rescaling quantities which appear in our final expressions by $R_5$. 

The tuples of partitions label various instanton configuration, and from the localization perspective the corresponding contributions are naturally expressed in terms of so-called arm-lengths and leg-lengths of boxes in these partitions \cite{Nek,Nek-Ok,NY-I,NY-II}. These expressions are summarized in the appendix \ref{app-partitions}. However these expressions can also be written purely in terms of the length of rows of these partitions, which is more useful from the perspective of putative matrix models we are after. As reviewed in the appendix \ref{app-partitions}, in this form the instanton partition functions for vector and (anti)fundamental multiplets read
\bea
Z^{4d} & = & \sum_{\vec{\lambda}=(\lambda^{(1)},\ldots,\lambda^{(n)})} \Lambda^{2n |\vec{\lambda}|} 
Z^{4d}_{\vec{\lambda},vec} \, Z^{4d}_{\vec{\lambda},(anti)fund}, \nonumber \\
Z^{4d}_{\vec{\lambda},vec} 
& = & \frac{1}{\epsilon_2^{2n |\vec{\lambda}|}}
\prod_{(l,i)\neq (k,j)} \frac{\Gamma(\lambda^{(l)}_i - \lambda^{(k)}_j +\beta(j - i) + b_{lk} + \beta)}{\Gamma(\lambda^{(l)}_i - \lambda^{(k)}_j +\beta(j - i) + b_{lk})} \frac{\Gamma(\beta(j - i) + b_{lk})}{\Gamma(\beta(j - i) + b_{lk} + \beta)}
 \label{Zinst4d}   \\
Z^{4d}_{\vec{\lambda},fund} 
&=&  \prod_{l=1}^n \prod_{\bf f=1}^{N_{\bf f}} \prod_{i=1} \frac{\Gamma(\lambda^{(l)}_i + b_l -M_{\bf f} -i\beta + 1)}{\Gamma(b_l -M_{\bf f} -i\beta + 1)}   \label{Zinst4dflav} 
\eea
where $|\vec{\lambda}|=\sum_{l=1}^n \sum_i |\lambda^{(l)}_i|$. The contribution to $Z^{4d}_{\vec{\lambda},antifund}$ is analogous as to $Z^{4d}_{\vec{\lambda},fund}$, but with $m_{\bf f}$ replaced by $\epsilon_1+\epsilon_2-m_{\bf f}$.

In the five-dimensional case we introduce
\be
q=e^{\epsilon_2}, \qquad \quad t=e^{-\epsilon_1}, \qquad \quad  Q_{l} = e^{a_l}\equiv e^{\epsilon_2 b_{l}}, \quad \qquad  Q_{l,k} = e^{a_l-a_k} \equiv e^{\epsilon_2 b_{lk}}.  \label{qtQ}
\ee
The partition functions read (again see the appendix \ref{app-partitions})
\bea
Z^{5d} & = & \sum_{\vec{\lambda}=(\lambda^{(1)},\ldots,\lambda^{(n)})} \Lambda^{2n |\vec{\lambda}|} 
Z^{5d}_{\vec{\lambda},vec}   \, Z^{5d}_{\vec{\lambda},(anti)fund} Z^{5dCS}_{\vec{\lambda},m_{CS}} ,  \nonumber  \\
Z^{5d}_{\vec{\lambda},vec}  
& = & \frac{1}{\epsilon_2^{2n |\vec{\lambda}|}}
\prod_{(l,i)\neq (k,j)} \frac{(Q_{l,k} q^{\lambda^{(l)}_i - \lambda^{(k)}_j} t^{j - i}; q)_{\infty}}{(Q_{l,k} q^{\lambda^{(l)}_i 
- \lambda^{(k)}_j} t^{j - i + 1}; q)_{\infty}}
 \frac{(Q_{l,k} t^{j - i +1}; q)_{\infty}}{(Q_{l,k} t^{j - i}; q)_{\infty}}
 \label{Zinst5d}   \\
Z^{5d}_{\vec{\lambda},fund} & = & \prod_{l=1}^n \prod_{\bf f=1}^{N_{\bf f}} \prod_{i=1} \frac{(q^{b_l -M_{\bf f} -i\beta + 1};q)_{\infty}}{(q^{\lambda^{(l)}_i + b_l -M_{\bf f} -i\beta + 1};q)_{\infty}}   \label{Zinst5dflav}  \\
 Z^{5dCS}_{\vec{\lambda},m_{CS}} & = & \prod_{l=1}^n Q_l^{-m_{CS} |\lambda^{(l)}|}  q^{\frac{-m_{CS} ||\lambda^{(l)}||^2}{2}} t^{\frac{m_{CS} ||\lambda^{(l),t}||^2}{2}}    \label{Z5dCS} 
\eea
where $||\lambda||=\sum_i \lambda_i^2$. The contributions for hypermultiplets in antifundamental representations $Z^{4d}_{\vec{\lambda},antifund}$ arise analogously as in the four-dimensional case. The term $Z^{5d}_{\vec{\lambda},m_{CS}}$ encodes Chern-Simons terms which may arise in the five-dimensional case, and which are parametrized by a single integer $m_{CS}$; such Chern-Simons contribution vanish in the four-dimensional limit. 

More generally, one can also consider contributions from matter in bifundamental or adjoint representations. The generalization of matrix models presented here to those cases is not difficult, and we leave it for future work. In principle, extending our methods to bifundamental matter would lead to multi-matrix models, while analysis of adjoint matter would generalize the results of \cite{matrix2star}.

We therefore focus on contributions for vector and (anti)fundamental multiplets, as well as Chern-Simons terms. To sum up, we note that the partition functions given above can be written, up to inessential constant factors, in a unified way as
\bea
Z & = & \sum_{\vec{\lambda}=(\lambda^{(1)},\ldots,\lambda^{(n)})} \Lambda^{2n |\vec{\lambda}|} 
Z_{vec}  \, Z_{(anti)fund} \,  Z^{CS}_{m_{CS}}, \nonumber  \\
Z_{vec} & = & \prod_{(l,i)\neq(k,j)} \frac{f \big(\lambda^{(l)}_i - \lambda^{(k)}_j +\beta(j - i) + b_{lk} + \beta \big)}{f \big(\lambda^{(l)}_i - \lambda^{(k)}_j +\beta(j - i) + b_{lk} \big)} \frac{f \big(b_{lk} + \beta(j - i) \big)}{f \big(b_{lk} + \beta(j - i) +  \beta \big)}   \label{Zf} \\
Z_{fund} & = & \prod_{l=1}^n \prod_{\bf f=1}^{N_{\bf f}} \prod_{i=1} \frac{f(\lambda^{(l)}_i + b_l -M_{\bf f} -i\beta + 1)}{f(b_l -M_{\bf f} -i\beta + 1)}  \label{Zf-fund}
\eea
where, respectively in the four- and five-dimensional case, the function $f(x)$ takes the form
\bea
f^{4d}(x) = \Gamma(x),\qquad \qquad f^{5d}(x) = \frac{1}{(q^x;q)_{\infty}}.     \label{f4d5d}
\eea
In five dimensions the Chern-Simons term $Z^{5dCS}_{m_{CS}}$ can also be expressed purely in terms of the lengths $\lambda^{(l)}_i$, which we will discuss in more detail in section \ref{sec-5d}. In the four-dimensional case we simply put $Z^{4dCS}_{m_{CS}} = 1$.


\section{Matrix model representation}         \label{sec-matrixmodel}

Nekrasov partition functions involve sums over $n$-tuples of partitions $\lambda^{(l)}$. In the previous section we expressed these sums purely in terms of the lengths of rows of these partitions $\lambda^{(l)}_i$. In this section we reinterpret these lengths as eigenvalues of matrices in certain ensembles. We follow the general strategy presented in \cite{SW-matrix,matrix2star,eynard-planch}. In this section we do not consider Chern-Simons terms in five-dimensional case, or assume that $m_{CS}=0$ in (\ref{Z5dCS}). Arbitrary $m_{CS}$ will be considered in section \ref{ssec-CS}.

\begin{figure}[htb]
\begin{center}
\includegraphics[width=\textwidth]{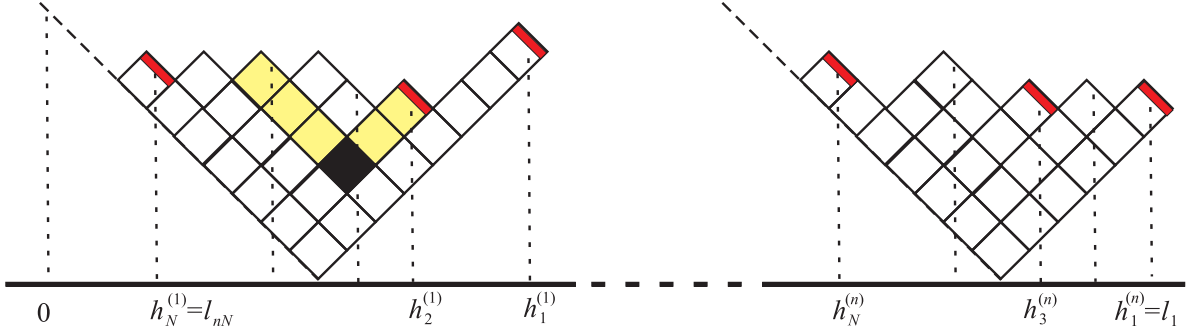} 
\begin{quote}
\caption{\emph{Translation of $\lambda^{(l)}_i$'s to $h_i^{(l)}$'s and $l_i$'s, which will be reinterpreted as matrix eigenvalues. The leg-length and arm-length of given box (in black) are defined respectively as the number of boxes to the right, and above this box (in yellow).}} \label{fig-hR}
\end{quote}
\end{center}
\end{figure}

To start with, we truncate the sums over partitions in Nekrasov partition functions, to sums over partitions with at most $N$ non-zero rows. After several manipulations we will find a matrix model representation of Nekrasov partition functions in the large $N$ limit of ensuing expressions. Having fixed $N$, we introduce
\be
h^{(l)}_i = \lambda^{(l)}_i - i + N + b_l,     \label{hl}
\ee
as shown in figure \ref{fig-hR}. In terms of these variables, the expressions (\ref{Zf}) and (\ref{Zf-fund}) take respectively the form
\bea
Z_{vec} & = & \prod_{(l,i)\neq(k,j)} \frac{f\big(h^{(l)}_i - h^{(k)}_j +(i-j)(1-\beta) + \beta \big)}{f \big(h^{(l)}_i - h^{(k)}_j +(i-j)(1-\beta) \big)} \frac{f \big(b_{lk} + \beta(j - i) \big)}{f\big(b_{lk} + \beta(j - i) + \beta \big)},
 \label{Zff}   \\
Z_{fund} & = & \prod_{l=1}^n \prod_{\bf f=1}^{N_{\bf f}} \prod_{i=1}^N \frac{f(h^{(l)}_i -M_{\bf f} -N + i(1-\beta) + 1)}{f(b_l -M_{\bf f} -i\beta + 1)} .     \label{Zff-fund}
\eea

The first important observation is that the above expressions can be written in terms of a single set of variables obtained from concatenation of sequences $h^{(l)}$
\be
l_{i=1,\ldots,nN} = (h^{(n)}_1,\ldots,h^{(n)}_N,\ldots \ldots,h^{(1)}_1,\ldots,h^{(1)}_N).
\ee
Assuming that $a_l$'s are decreasing and in sufficiently large distances from each other, the sequence $l_{i=1,\ldots,nN}$ is decreasing. As shown in appendix \ref{app-sums}, the expressions (\ref{Zff}) and (\ref{Zff-fund}), for arbitrary $f$ (not necessarily of the form (\ref{f4d5d})), can now be written as
\bea
Z_{vec} & = & \Big( \frac{f(\beta)}{f(0)} \Big)^{nN} \,  \prod_{i \neq j}^{nN} \frac{f\big(l_i - l_j +(i_{mod\, N}-j_{mod\, N})(1-\beta) + \beta \big)}{f \big(l_i - l_j +(i_{mod\, N}-j_{mod\, N})(1-\beta) \big)}   \times \label{Zff-Vand} \\
& & \qquad  \times \prod_{l=1}^n  \prod_{i=1}^{nN}  \frac{f\big(-(l_i - b_l + (i_{mod\, N}-N)(1-\beta)) \big)}{f\big(l_i - b_l + (i_{mod\, N}-N)(1-\beta) + \beta \big)},     \label{Zff-l}   \\
Z_{fund} & = & \prod_{\bf f=1}^{N_{\bf f}} \prod_{i=1}^{nN} \frac{f(l_i -M_{\bf f} -N + i_{mod\, N}(1-\beta) + 1)}{f(b_l -M_{\bf f} -i_{mod\, N}\beta + 1)}  \label{Zff-fund-l}
\eea
where $i_{mod\, N}\equiv i\, mod\, N$. Moreover, as $l_i$ are expected to be large, we introduce rescaled variables 
\be 
x_i=\epsilon_2 l_i.    \label{xl}
\ee

Now we wish to reinterpret the above expressions as partition functions of $\beta$-deformed matrix models, with $l_i$'s playing the role of eigenvalues. As we show below, the functions $f(x)$ indeed have a form relevant for such an interpretation, at least in the leading expansion in $\epsilon_2$. As we will see, the subleading terms in such an expansion depend not only on variables $l_i$, but also $i\,mod\, N$. This makes matrix interpretation of those subleading terms less clear, however, as explained before, they should be inessential to the solution of our matrix models. We also note, that such dependence on $i$ in the subleading terms does not arise in the $\beta=1$ case, see also \cite{SW-matrix,matrix2star}. 

To relate the above expressions to matrix models, in what follows we will perform the following steps:
\begin{enumerate}
\item Reinterpret the factor (\ref{Zff-Vand}) in $Z_{\vec{\lambda},vec}$ involving differences $(l_i-l_j)=(x_i-x_j)/\epsilon_2$ as the (possibly deformed) Vandermonde determinant
\item Reinterpret other factors (\ref{Zff-l}), (\ref{Zff-fund-l}), etc., involving just a single $l_i=x_i/\epsilon_2$, as contributions to the potentials $V_{vec}, V_{fund}$, etc., by writing them in the form
\be
\exp \Big[ {-\frac{1}{\epsilon_2}\sum_{i=1}^{nN}\big(V_{vec}(x_i) + V_{fund}(x_i) + \ldots} \big) \Big]   \label{eTrV}
\ee
\item Introduce an auxiliary function $f_{poles}(x)$, which has simple poles at all integer values of the argument
\item Replace the summation over sequences of $l_i$ by the sum over all $l_i$ using symmetrization, and subsequently by the integration over $x_i=\epsilon_2 l_i$ over a contour $C$ which encircles a (part of) real axis:
\be
\sum_{l_1>l_2>\ldots>l_{nN}}\ldots \quad \longrightarrow \quad \frac{1}{(nN) !} \sum_{l_1, l_2, \ldots, l_{nN}}\ldots \quad \longrightarrow \quad  \frac{1}{(nN) !} \oint_C d^{\, nN} x \,\prod_{i=1}^{nN} f_{poles}(x_i)\ldots \label{sumint}
\ee
\end{enumerate}
Then, in the leading expansion in $\epsilon_2$, the resulting expressions have the form of the eigenvalue representation of matrix integrals. In the following sections we discuss this program separately for four- and five-dimensional theories.


\section{Matrix models for four-dimensional theories}               \label{sec-4d}

In this section we reinterpret four-dimensional Nekrasov partition functions as matrix models in the leading expansion in $\epsilon_2$, and discuss the example of $SU(2)$ theory with two fundamentals and two antifundamentals.

\subsection{Derivation of matrix models}

Here we follow the steps $1-4$ listed in the previous section, and apply them to four-dimensional Nekrasov partition functions.

\begin{enumerate}
\item The factor (\ref{Zff-Vand}) for the four-dimensional theory is realized in terms of $f^{4d}(x)=\Gamma(x)$ functions (\ref{f4d5d}). This factor is therefore a ratio of gamma functions. Here comes the crucial observation: the relation to the $\beta$-Vandermonde arises from the asymptotics of the ratio of gamma functions (\ref{asympt-Gamma}) presented in appendix \ref{app-ratios}. As we expect $l_i,l_j$ to be large, we can make the following identification between (\ref{asympt-Gamma}) and the arguments of $f^{4d}$
\be
z = l_i-l_j,  \qquad \alpha = (i_{mod\, N} - j_{mod\, N})(1-\beta),\qquad \beta\equiv \beta.      \label{zalphabeta}
\ee
Therefore, expressing this expansion in terms $x_i$ introduced in (\ref{xl}), we get
\be
\prod_{i \neq j}^{nN} \frac{\Gamma\big(l_i - l_j +(i_{mod\, N}-j_{mod\, N})(1-\beta) + \beta \big)}{\Gamma\big(l_i - l_j +(i_{mod\, N}-j_{mod\, N})(1-\beta) \big)}  = \epsilon_2^{-\beta n N} \prod_{i \neq j}^{nN}  (x_i - x_j)^{\beta} \Big(1 + \mathcal{O}(\frac{\epsilon_2}{x_i-x_j}) \Big).
\ee
As presented in appendix \ref{app-ratios}, the subleading terms are of the form
$$
\mathcal{O}(\frac{\epsilon_2}{x_i-x_j}) = \sum_{k=1}^{\infty} C_k(\beta,\alpha) \frac{\epsilon_2^k}{(x_i-x_j)^k}.
$$
We note that the dependence on the coefficient $\alpha = (i_{mod\, N} - j_{mod\, N})(1-\beta)$ is encoded only in higher order coefficients $C_n(\beta,\alpha)$. Note that these coefficients vanish in the $\beta=1$ case. This is why no $\hbar$-deformation of the measure was observed in \cite{SW-matrix,matrix2star}.

\item Using the expansion of the logarithm of the gamma function (\ref{lnGamma}), the remaining factors from $Z^{4d}_{vec}$, given in (\ref{Zff-l}) with $f^{4d}(x)=\Gamma(x)$, can be interpreted as contributions to the potential in (\ref{eTrV}). The full form of $V^{4d}_{vec}(x)$ is given in (\ref{V4dvec-full}). To the leading order in $\epsilon_2$ it reads
\be
V^{4d}_{vec}(x) = 2 \sum_{l=1}^n \Big[(x-a_l)\log(x-a_l) -(x-a_l) + \mathcal{O}(\epsilon_2) \Big].            \label{V4dvec-leading}
\ee
The same potentials were obtained in the leading order in the $\beta=1$ case in \cite{SW-matrix}. As explained earlier, such leading contributions should be sufficient to get the spectral curve of the matrix model, and subsequently solve it. In fact the subleading terms, for general $\beta$, do not depend only on $\lambda^{(l)}_i$, but also explicitly on $(i\,mod\, N)$, therefore they cannot be simply symmetrized and their matrix model interpretation is less clear. Nonetheless, as an example we present the full form of $V^{4d}_{vec}(x)$ in appendix \ref{app-V4dvec}. Similarly as above, we also get the contribution for fundamental multiplets
\be
V^{4d}_{fund}(x;m) =  \sum_{f=1}^{N_f} \Big[-(x-m_{\bf f})\log(x-m_{\bf f}) + (x-m_{\bf f}) + \mathcal{O}(\epsilon_2) \Big],            \label{V4dfund-leading}  
\ee
and we omit the constant factors of the form $\prod_{l,i,f}\Gamma(b_l-M_{f}-i\beta+1)$ which can be absorbed into the normalization of $Z$. 
The contribution for antifundamentals is obtained by the substitution $m_{\bf f}\to \epsilon_1+\epsilon_2-m_{\bf f}$.
We note that the potentials for the fundamental matter include Penner-like factors of the form $m_{\bf f}\log(x-m_{\bf f})$ which appeared in the work of Dijkgraaf and Vafa \cite{DV2009}. Such terms also arise in matrix models for 2* theories analyzed in \cite{matrix2star}.

\item As in \cite{SW-matrix,matrix2star,eynard-planch}, in the four-dimensional case we introduce the function
$$
f^{4d}_{poles}(x) = -x \G(-x) \G(x) e^{-i \pi x} = \frac{\pi e^{-i\pi x}}{\sin(\pi x)},
$$
with simple poles at all integer values of its argument. 
Upon integration along the contour $C$ encircling $[ b_1,\infty [$ 
part of the real axis, this function can be used to pick up all 
integer values of $h^{(l)}_i\in [ b_1,\infty [$.

\item Finally we replace the summation over $l_i$ by the integration over $x_i$ according to (\ref{sumint}). This leads to the following expression
\be
Z^{4d} =  \int_{Mat_{nN}} \mathcal{D}M e^{-\frac{1}{\epsilon_2} \Tr V^{4d}(M)},   \qquad \qquad \mathcal{D}M=\prod_i dx_i \prod_{i<j}(x_i-x_j)^{2\beta},  \label{Z4d-all}
\ee
where $M\in Mat_{nN}$ represents $nN\times nN$ matrix form the $\beta$-ensemble, and the measure $\mathcal{D}M$ involves the $\beta$-deformed Vandermonde determinant in the leading order in $\epsilon_2$. Similarly, to the leading order the potential is given by
$$
V^{4d}(x) = tx + V^{4d}_{vec}(x) + V^{4d}_{fund}(x;m)
$$
where we included the linear term arising from the instanton counting parameter $\Lambda$. The contributions from vector and fundamental multiplets are given respectively in (\ref{V4dvec-leading}) and (\ref{V4dfund-leading}).
The example of a potential for pure $SU(2)$ theory is given in figure \ref{fig-SU2}.

\begin{figure}[htb]
\begin{center}
\includegraphics[width=0.5\textwidth]{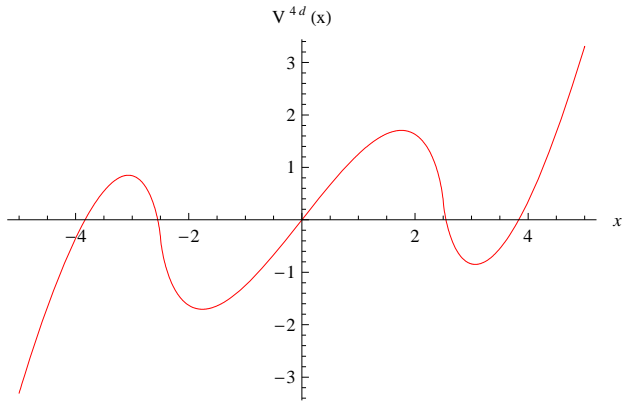} 
\begin{quote}
\caption{\emph{Matrix model potential for the pure $SU(2)$ Seiberg-Witten theory.}} \label{fig-SU2}
\end{quote}
\end{center}
\end{figure}

\end{enumerate}


\subsection{Example -- $SU(2)$ theory with 4 hypermultiplets}

$SU(2)$ theory with 4 hypermultiplets is a simple and important example relevant for the AGT correspondence. Let us denote masses of two fundamental multiplets by $m_1,m_2$, and two antifundamental ones by $-m_3,-m_4$. We also put $a_1=-a_2=a$. Then the matrix model for this theory is given by (\ref{Z4d-all}), with the potential which can be written as
$$
V^{4d}_{SU(2),N_{\bf f}=4} = x\log\frac{(x-a)^2(x+a)^2}{(x-m_1)(x-m_2)(x-m_3)(x-m_4)}  - a\log \frac{x-a}{x+a} + 
$$
$$
+ \sum_{{\bf f}=1}^4 m_{\bf f}\log(x-m_{\bf f}).
$$
The logarithmic Penner-like terms in the second line appear similarly in \cite{DV2009}. As discussed in \cite{AGT}, to obtain the instanton contribution to the $SU(n)$ theory, apart from setting $a_1=-a_2$ one still has to isolate the appropriate $U(1)$ factor. It would be interesting to understand if such factors play any role from the matrix model perspective. Nonetheless, we suppose we can treat them as an overall contribution, without changing the form of the matrix model.


\section{Matrix models for five-dimensional theories}         \label{sec-5d}
 
In this section we derive matrix models for five-dimensional theories. To start with, we follow the steps $1-4$ from the previous section, in case with $m_{CS}=0$. Next we discuss arbitrary $m_{CS}$, which corresponds to turning on Chern-Simons terms.

\subsection{Theories with fundamental matter}

Here we again follow the steps $1-4$ listed in the previous section, now in the context of five-dimensional theories.

\begin{enumerate}
\item The factor (\ref{Zff-Vand}) for four-dimensional theory is realized in terms of $f^{5d}(x)=(q^x;q)^{-1}_{\infty}$ functions (\ref{f4d5d}). With analogous identification as in (\ref{zalphabeta}), expressing the result in terms of $x_i$ variables (\ref{xl}), and using the asymptotics (\ref{asympt-qGamma}) we get
\be
\prod_{i \neq j}^{nN} \frac{\big(q^{l_i - l_j +(i_{mod\, N}-j_{mod\, N})(1-\beta)};q \big)_{\infty}}{\big(q^{l_i - l_j +(i_{mod\, N}-j_{mod\, N})(1-\beta) + \beta};q\big)_{\infty}}  \simeq \prod_{i \neq j}^{nN}  (1 - e^{x_i - x_j})^{\beta}.      \label{Vandermonde5d}
\ee
Up to the overall phase factor, this can be written as
\be
\prod_{i < j}^{nN}  \Big(2\sinh\frac{x_i - x_j}{2}\Big)^{2\beta}.              \label{Vandermonde-sinh}
\ee

\item The remaining factors from $Z^{5d}_{vec}$, given in (\ref{Zff-l}) with $f^{5d}(x)=(q^x;q)^{-1}_{\infty}$, can be written as in (\ref{eTrV}). Using the notation of the quantum dilogarithm (\ref{g-dilog}), and then the asymptotics (\ref{g-dilog-asymp}), this leads to the expression
$$
\prod_{l=1}^n \prod_{i=1}^{nN} \frac{g(q^{-l_i+b_l -(i_{mod\, N} - N)(1 - \beta)})}{g(q^{l_i-b_l +(i_{mod\, N} - N)(1 - \beta) + 1})} \simeq \prod_{l=1}^n \prod_{i=1}^{nN} e^{\frac{1}{\epsilon_2} \big( \Li_2(e^{x_i-a_l}) - \Li_2(e^{-x_i+a_l}) \big)  }
$$
where $x_i=\epsilon_2 l_2$. From the inversion relation for the dilogarithm
$$
\Li_2(z) + \Li_2(z^{-1}) = -\frac{1}{2}(\log z)^2 + \frac{\pi^2}{3} -i\pi\log z,
$$
we get, for the $U(n)$ theory, up to constant and imaginary factors and to the leading order in $\epsilon_2$ 
\be
V^{5d}_{vec}(x) = \frac{n}{2}x^2 + 2\sum_{l=1}^n \Li_2\big(e^{-x+a_l}\big).                 \label{V5dvec-leading}
\ee
This potential agrees with results of \cite{SW-matrix}.\footnote{The opposite sign of $x$ in the exponent appears due to a different convention $q=e^{\epsilon_2}$ here, versus $q=e^{-g_s}$ in \cite{SW-matrix}.} In a similar way, the contributions from fundamental multiplets (\ref{Zff-fund-l}) give
\be
V^{5d}_{fund}(x) = \sum_{\bf f} \Li_2\big(e^{x-m_{\bf f}}\big).                 \label{V5dfund-leading}
\ee

\item The quantum dilogarithm (\ref{g-dilog}) vanishes, $g(q^h)=0$, for $h$ a positive integer. At such points its derivative is
$$ 
g'(q^h) = - \frac{g(1)^2 e^{i\pi h} q^{-h(h-1)/2}}{q^h(1-q^h)g(q^{-h})}.
$$
Therefore the following function has simple poles with residue 1 for $x=q^h$ with $h\in\mathbb{N}$
$$
f^{5d}_{poles}(x) = -\frac{g(1)^2 e^{-\frac{i\pi}{g_s}\log x} e^{\frac{(\log x)^2}{2g_s}}}{(1-x)\sqrt{x}g(x)g(x^{-1})}.
$$
Written in the exponential form, it contributes only a linear term in $x$ to the potential \cite{SW-matrix,eynard-planch}.

\item Finally we get
\be
Z^{5d} = \int_{Mat_{nN}}  \mathcal{D}M e^{-\frac{1}{g_s} \Tr V^{5d}(M)},\qquad\qquad  \mathcal{D}M = \prod_i d x_i  \prod_{i<j} \Big(2 \sinh \frac{x_i - x_j}{2}\Big)^{2\beta}.
\ee
Now the measure is given by the Vandermonde determinant which is both $\beta$-deformed and sinh-deformed, while the potential reads
$$
V^{5d}(u) = tx + V^{5d}_{vec}(x) + V^{5d}_{fund}(x), 
$$
where we again included the linear term arising from instanton counting parameter $\Lambda$, as well as $f^{5d}_{poles}(x)$. The contributions for vector multiplet and hypermultiplet are given respectively in (\ref{V5dvec-leading}) and (\ref{V5dfund-leading}).
\end{enumerate}


\subsection{Chern-Simons terms and more general Calabi-Yau manifolds}     \label{ssec-CS}

In five-dimensional theories one can also include Chern-Simons terms \cite{Intriligator:1997pq,Tachikawa} of the form
$$
\int c_{ijk} \, A_i\wedge F_j\wedge F_k,
$$
with indices labeling vector multiplets. It turns out that there is only a discrete consistent choice of $c_{ijk}$ and they are labeled by a single integer $m_{CS}$. Five-dimensional theories with these Chern-Simons terms are also equivalent to topological string theory on appropriate toric Calabi-Yau manifolds \cite{geom-eng}. In this context the constants $c_{ijk}$ translate to triple intersection numbers of these manifolds. This dictionary was discussed in detail in \cite{SW-matrix}. 

The equivalence between Nekrasov partition functions, with $\epsilon_1=-\epsilon_2=\hbar$, and topological string amplitudes computed from the topological vertex has been explicitly checked in \cite{SUNamplitudes}. It turns out that this equivalence extends also to the case of arbitrary $\epsilon_1,\epsilon_2$. It was proposed that topological string amplitudes in this case should be refined to two parameters $(q=e^{\epsilon_2},t=e^{-\epsilon_1})$ in terms of the refined topological vertices \cite{refined-vertex,refined-Kanno}. In \cite{Taki,AwataKanno-refined} it has been checked that such refined amplitudes indeed reproduce Nekrasov partition functions and the results for both vertices agree, and the consistent refinement of the Chern-Simons terms has also been found. It was shown that the refined Chern-Simons terms take form \cite{Taki,AwataKanno-refined}
$$
Z^{5dCS}_{\vec{\lambda},m_{CS}}(q,t) =  \prod_{l=1}^n \Big( Q_l^{|\lambda^{(l)}|}  q^{\frac{||\lambda^{(l)}||^2}{2}} t^{-\frac{||\lambda^{(l),t}||^2}{2}} \Big)^{-m_{CS}}, 
$$
where $||\lambda||=\sum_i \lambda_i^2$. This can be written explicitly in terms of the lengths of rows
\bea
Z^{5dCS}_{\vec{\lambda},m_{CS}}(q,t) & = & \prod_{l=1}^n \Big( Q_l^{|\lambda^{(l)}|} \Big(\frac{q}{t}\Big)^{\frac{||\lambda^{(l)}||^2}{2}} t^{\frac{1}{2}\kappa_{\lambda^{(l)}}} \Big)^{-m_{CS}} = \nonumber \\
& = & \prod_{l=1}^n Q_l^{-m_{CS} |\lambda^{(l)}|} e^{-\frac{1}{2}m_{CS} \epsilon_2\sum_i \big((\lambda^{(l)}_i)^2  + \beta \lambda^{(l)}_i (1-2i) \big)}  \label{CSterms}
\eea
where $\kappa_{\lambda} = \lambda_i(\lambda_i+1-2i) = ||\lambda||^2 - ||\lambda^t||^2$. Therefore, this expression can be written in terms of $h^{(l)}_i$ using (\ref{hl}).

Nonetheless, the resulting expression cannot be interpreted as a one-matrix model for non-zero $m_{CS}$. This is so, because each set of $h^{(l)}_i$ is coupled to different $a_l$ through the term $Q_l^{-m_{CS} |\lambda^{(l)}|}$. Therefore the symmetrization (\ref{sumint}) cannot be performed for the entire set of $l_i$'s, but only within each set $l_{(k-1)N+1},\ldots,l_{kN}$. In consequence we obtain the $n$-matrix model, with $n$ sets of eigenvalues $x^{(l)}_i=\epsilon_2 h^{(l)}_i$, and with the leading contribution to the potential of the linear form
\be
V^{5dCS} = m_{CS}\sum_{l=1}^n \Big(a_l \sum_{i=1}^N x^{(l)}_i \Big).     \label{VCS}
\ee
We also note, that while it is straightforward to include the quadratic terms appearing in the last exponent in (\ref{CSterms}), they cannot be reinterpreted as the contribution to the potential of a matrix model due to an explicit dependence on $i$. Happily these terms are subleading in $\epsilon_2$. Therefore (\ref{VCS}) is the only leading contribution to the matrix model potential $V^{5dCS}$, and this is the same as in $\epsilon_1=-\epsilon_2=\hbar$ case discussed in \cite{SW-matrix}.


\subsection{Relation to other deformations}     \label{ssec-deform}

We found that in five dimensional theories, in the leading $\epsilon_2$ expansion, the Vandermonde determinant takes a form of a sinh-deformation (\ref{Vandermonde-sinh}), familiar from \cite{HIV,SW-matrix,matrix2star,eynard-planch}. In the context of matrix models related to Nekrasov partition functions, another -- seemingly unrelated  -- deformation was postulated in \cite{SchiappaWyllard}, which amounts to replacing the $\beta$-deformed Vandermonde determinant by 
\be
\hat{V}^q_{\beta} = \prod_{I<J}^N \lambda_I^{2\beta} \frac{(\frac{\lambda_I}{\lambda_J}q^{-\beta};q)_{\infty}}{(\frac{\lambda_I}{\lambda_J}q^{\beta};q)_{\infty}}.    \label{Vand-SchWyll}
\ee
The identification $\lambda_I=q^{l_I}$ makes contact with our notation.
As discussed in \cite{SchiappaWyllard} the integral\footnote{Note a different sign convention $\epsilon_{1,2} \to \varepsilon_{1,2}=-\epsilon_{1,2}$ in \cite{SchiappaWyllard}.}
$$
S^q(\alpha_1,\alpha_2,\beta;z)=\int \prod_{I=1}^N d_q \lambda_I \, \hat{V}^q_{\beta} \, (z-\lambda_I q^{-1/2}) \lambda_I^{-2\alpha_I/\epsilon_1} \frac{(\lambda_I q^{\alpha_2/\epsilon_1};q)_{\infty}}{(\lambda_I q^{-\alpha_2/\epsilon_1};q)_{\infty}},
$$ 
which includes this supposed deformed determinant, reproduces Nekrasov partition function for $SU(2)$ theory with four fundamentals, with a special choice of parameters $\alpha_3=-\epsilon_1/2$. This expression is in fact a Jackson integral, i.e. a discrete integral defined by $\int d_q x f(x) = (1-q)\sum_{k=0}^{\infty} f(q^k)q^k$, which reproduces Riemann integral in $q\to 1$ limit. The value of the above integral can be expressed in terms of $q$-deformed Jack or Jacobi polynomials.

We now note several similarities between these expressions and matrix models discussed in this paper. Firstly, the discreteness in variables $\lambda_I$ in the Jackson integral is of similar kind as the discreteness of eigenvalues in our matrix models discussed in previous sections. Secondly, we observe that the deformation (\ref{Vand-SchWyll}) is closely related to the determinant-like expression arising explicitly from rewriting Nekrasov partition function, given in (\ref{Vandermonde5d}). Indeed, both expressions differ by terms of the form $(i_{mod\, N}-j_{mod\, N})(1-\beta)$ in (\ref{Vandermonde5d}), which from our perspective get multiplied by $\epsilon_2$ and are subleading. Up these terms, the form of both expressions in the asymptotic form (\ref{asympt-qGamma}) is the same, and in particular they both lead to the sinh-deformed Vandermonde (\ref{Vandermonde-sinh}) in the leading order. The insertions of $(\lambda_I q^{\pm \alpha_2/\epsilon_1};q)_{\infty}$ in the integral $S^q(\alpha_1,\alpha_2,\beta;z)$ are given by the same infinite products as contributions from vector- and fundamental multiplets in (\ref{Zff-l}) and (\ref{Zff-fund-l}) with $f^{5d}(x)=(x;q)^{-1}_{\infty}$. This suggests, that one might indeed express matrix models for Nekrasov partition functions, to all orders in $\epsilon_2$ deformation, in terms of Jack or Jacobi polynomials. 

Moreover, it was suggested in \cite{SchiappaWyllard} that the deformation of the measure to the five-dimensional case could be related to the $q$-deformed Virasoro algebra considered in \cite{qVirAGT}. If this is indeed the case, it would be interesting to see how such an algebra manifests on the level of our matrix models. For example, in the standard formulation of matrix models, the loop equations are equivalent to Virasoro constraints. These constraints can be written in terms of operators which satisfy the Virasoro algebra and annihilate the matrix model partition function. It is interesting whether a similar structure arises in the $q$-deformed case. It is also tempting to extend such a deformation of Virasoro algebra to the six-dimensional case.

Certain deformations of Virasoro algebra are also related to the so-called $Q$-bosons and $Q$-fermions, and $Q$-deformed boson-fermion correspondence. These objects also arise in the context of topological strings \cite{deform}. Even though in this case the role of the deformation is different, i.e. it plays a role of the K{\"a}hler parameter, it would be interesting to study if there are some relations between these both deformations of the Virasoro algebra. 




\section{Further research}  \label{sec-discussion}

In this paper we derived matrix models for $\beta$-ensembles, which encode the instanton part of Nekrasov partition functions. This is just the first of several steps which should be completed. Most of all, it is important to analyze these models using matrix model techniques and confirm the relation to Seiberg-Witten theories from this perspective. Typical matrix model analysis involves finding the spectral curve and solving the loop equations. In the present case this is entirely non-trivial: as shown in \cite{beta-matrix-1,beta-matrix-2}, matrix models for $\beta$-ensembles lead to non-commutative spectral curves. Moreover the form of our potentials is more complicated than the polynomial potentials analyzed in \cite{beta-matrix-1,beta-matrix-2}, and therefore the entire theory has still to be extended. Apart from general interest, the ultimate goal of this program, as explained in the introduction, would be to provide more rigorous proof of the AGT conjecture.

Even though our models share several similarities (deformed measures, Penner-like potentials) with matrix models discussed in \cite{DV2009,PennerEguchi,SchiappaWyllard}, there are also some differences. Apart from the Penner-like terms, our potentials include additional contributions. Some parameters arise on different footing: for example the instanton counting parameter $\Lambda$ is encoded in the linear term in our models, while in \cite{DV2009} is appears under the logarithm. It would be important to elucidate these discrepancies. It is also important to extend our analysis to bifundamental fields and quiver gauge theories, and correspondingly Toda systems. Matrix models for such six-dimensional theories, with elliptic deformations, can also be written down explicitly.

It would also be interesting to reexpress our results as matrix models in terms of Jack or Jacobi polynomials. Supposedly, the issue of symmetrization of the subleading terms should not arise in this case. Nonetheless, to the leading order such putative matrix models must agree with ours, and they should lead to the same (non-commutative) spectral curve and the entire solution.

Assuming that the dualities discussed in the introduction indeed hold, it is important to understand how various features of gauge theories and Liouville theories manifest in terms of our matrix models. One should understand the role of the one-loop part of the Nekrasov partition functions, as well as the decoupling of the $U(1)$ factor, which play an important role in the AGT conjecture. Some other such features include: the relation to the $q$-deformed Virasoro algebra (discussed briefly in section \ref{ssec-deform}), extension of this relation to the six-dimensional elliptic case (and finding appropriate more general deformation of the Virasoro algebra), interpretation of recursion relations between conformal blocks, matrix model interpretation of surface and loop operators discussed in \cite{AGGTV-surface,DGOT-loop}, and many others.

Finally, it is desirable to elucidate physical and mathematical interpretation of our matrix models. Physically, it is tempting to connect our results to phenomena involving D-branes and geometric transitions, underlying the whole Dijkgraaf-Vafa program \cite{DV-0206,pert-window,DV2009}. Mathematically, our results hint into new structures hidden in instanton moduli spaces, which underlie the localization program of Nekrasov \cite{Nek,Nek-Ok,NY-I,NY-II}.


\newpage

\appendix

\section{More on Nekrasov partition functions}      \label{app-partitions}

In general, instanton parts of Nekrasov partition functions take the form 
$$
Z = \sum_{\vec{\lambda}=(\lambda^{(1)},\ldots,\lambda^{(n)})} \Lambda^{2n |\vec{\lambda}|} 
Z_{\vec{\lambda},vec} \, Z_{\vec{\lambda},(anti)fund} \, Z_{\vec{\lambda},adj}\, Z_{\vec{\lambda},bifund} \,Z^{CS}_{\vec{\lambda},m_{CS}}, \nonumber \\
$$
where the summation extends over $n$ sets of two-dimensional partitions $\lambda^{(l)}$, $|\vec{\lambda}|=\sum_{l,i} \lambda^{(l)}_i$ is the total number of boxes in a given set of such partitions, and $\Lambda$ represents appropriate instanton counting parameter (i.e. dynamically generated scale in asymptotically free theories, or appropriately renormalized bare coupling in conformal theories). $Z_{\vec{\lambda},vec}$, $Z_{\vec{\lambda},(anti)fund}$, $Z_{\vec{\lambda},adj}$, $Z_{\vec{\lambda},bifund}$ and $Z^{CS}_{\vec{\lambda},m_{CS}}$ represent respectively contributions from vector multiplets, hypermultiplets in the (antifundamental or) fundamental representation, hypermultiplets in the adjoint representation, in the bifundamental representation, and (trivial in four-dimensional theories) Chern-Simons terms. These terms depend on the Coulomb branch parameters $a_l$ and masses of matter fields $m_{\bf f}$. The form of these terms depends also on the gauge group \cite{ABCD}. In this paper we focus on the $U(n)$ gauge groups, so we write below explicit expressions only in this case. The form of all these terms depends also on the dimensionality of the spacetime: five- and six dimensional terms are respectively trigonometric and elliptic generalizations of the four-dimensional ones \cite{NY-I,NY-II,HIV}. The contribution for antifundamental hypermultiplets $Z_{\vec{\lambda},antifund}$ is related to that of fundamental ones $Z_{\vec{\lambda},fund}$ by replacement of $m_{\bf f}$ by $\epsilon_1+\epsilon_2-m_{\bf f}$, so usually we write down explicitly only the latter ones. The contribution from adjoint or bifundamental hypermultiplets is not considered to much extent in this paper. To find the total partition function one has to insert into the above sum the appropriate number of these terms, corresponding to the field contents of the theory of interest. 

There are several representations of Nekrasov partition functions which are somehow scattered through literature, see e.g. \cite{Nek,Nek-Ok,NY-I,NY-II,AwataKanno-refined,Taki,AGT}. We discuss them briefly below. The representation which is most useful to construct our matrix models, discussed as the third one, appears to be used the least often. 

Firstly, various factors $Z_{\vec{\lambda},\ldots}$ introduced above can be written as product over all boxes in partitions $\lambda^{(l)}$. In case of vector, bifundamental or adjoint multiplets these products involve the arm-length $a_{\lambda^{(l)}}(\square)=\lambda^{(l)}_i - j$ and the leg-length $l_{\lambda^{(l)}}(\square)=\lambda^{(l),t}_j - i$ (shown in figure \ref{fig-hR}) of each box $\square=(i,j)\in\lambda^{(l)}$. The explicit form of these expressions is nicely summarized e.g. in \cite{AGT}. As they are not very useful for our purposes, we just just recall as an example that the contribution from fundamental hypermultiplets in the four-dimensional case takes the form \cite{AGT}
$$
Z^{4d}_{\vec{\lambda},fund}  =  \prod_{l=1}^n \prod_{\square=(i,j)\in \lambda^{(l)}} \big(a_{l} +i\epsilon_1 + j \epsilon_2 - m \big),  
$$
while vector multiplets in the five-dimensional case contribute \cite{AwataKanno-refined}
$$
Z^{5d}_{\vec{\lambda},vec}  =  \prod_{l,k=1}^n \frac{1}{\prod_{\square\in \lambda^{(l)}} \big(1 - q^{a_{\lambda^{(l)}}(\square)} t^{l_{\lambda^{(l)}}(\square) +1} Q_{l,k} \big)    \prod_{\square\in \lambda^{(k)}} \big(1 - q^{-a_{\lambda^{(k)}}(\square) - 1} 
t^{-l_{\lambda^{(k)}}(\square)} Q_{l,k} \big)}.
$$

The second form of Nekrasov partition functions involves lengths of rows of partitions $\lambda^{(l)}$, as well as transposed partitions $\lambda^{(l),t}$. In particular, the contribution for vector multiplets reads \cite{Nek-Ok}
$$
Z^{4d}_{\vec{\lambda},vec}  = 
\prod_{l,k;i,j} \frac{a_l-a_k + \epsilon_1(i-1) + \epsilon_2(-j)}{a_l-a_k + \epsilon_1(i - \lambda^{(k),t}_j-1) + \epsilon_2(\lambda^{(l)}_i-j)},
$$
while several equivalent forms of the five-dimensional contribution can be found in \cite{AwataKanno-refined}. The contribution for Chern-Simons terms in five-dimensional theories (discussed in section \ref{ssec-CS}) reads
$$
Z^{5dCS}_{\vec{\lambda},m_{CS}} = \prod_{l=1}^n Q_l^{-m_{CS} |\lambda^{(l)}|}  q^{\frac{-m_{CS} ||\lambda^{(l)}||^2}{2}} t^{\frac{m_{CS} ||\lambda^{(l),t}||^2}{2}},
$$
and this is trivial in the four-dimensional case.

The third form of Nekrasov partition functions involves just lengths of rows $\lambda^{(l)}_i$ and their differences. In such a form, the four-dimensional contributions for vector and fundamental multiplets read \cite{Nek-Ok}
\bea
Z^{4d}_{\vec{\lambda},vec}
& = & \frac{1}{\epsilon_2^{2n |\vec{\lambda}|}}
\prod_{(l,i)\neq (k,j)} \frac{\Gamma(\lambda^{(l)}_i - \lambda^{(k)}_j +\beta(j - i) + b_{lk} + \beta)}{\Gamma(\lambda^{(l)}_i - \lambda^{(k)}_j +\beta(j - i) + b_{lk})} \frac{\Gamma(\beta(j - i) + b_{lk})}{\Gamma(\beta(j - i) + b_{lk} + \beta)}
   \\
Z^{4d}_{\vec{\lambda},fund} & = &  
=  \prod_{l=1}^n \prod_{i=1} \frac{\Gamma(\lambda^{(l)}_i + b_l -M -i\beta + 1)}{\Gamma(b_l -M -i\beta + 1)},  
\eea
while those in five dimensions \cite{AwataKanno-refined,SchiappaWyllard}
\bea
Z^{5d}_{\vec{\lambda},vec} & = & 
\prod_{(l,i)\neq (k,j)} \frac{(Q_{l,k} q^{\lambda^{(l)}_i - \lambda^{(k)}_j} t^{j - i}; q)_{\infty}}{(Q_{l,k} q^{\lambda^{(l)}_i 
- \lambda^{(k)}_j} t^{j - i + 1}; q)_{\infty}}
 \frac{(Q_{l,k} t^{j - i +1}; q)_{\infty}}{(Q_{l,k} t^{j - i}; q)_{\infty}}
   \\
 Z^{5d}_{\vec{\lambda},fund} & = & \prod_{l=1}^n \prod_{\bf f=1}^{N_{\bf f}} \prod_{i=1} \frac{(q^{b_l -M_{\bf f} -i\beta + 1};q)_{\infty}}{(q^{\lambda^{(l)}_i + b_l -M_{\bf f} -i\beta + 1};q)_{\infty}}  
\eea
The form of the five-dimensional terms is in fact subtle, and depends on the operator to which the localization formula is applied. The more often encountered form involving sinh functions arises in case of the computation of the index of the Dirac operator, while the above form would correspond to the index of the Dolbeault operator. We find the above form more convenient, and the difference between the formulas involving sinh functions amounts simply to a redefinition of the linear term in the potential of matrix models which we find. From the geometric engineering viewpoint one could in fact consider various blow-ups representing matter multiplets, however we restrict only to the simplest case given above.


\section{Rewriting the sums}     \label{app-sums}

For an arbitrary function $f$, we wish to rewrite the contributions from the vector multiplet, given in (\ref{Zff})
\be
Z_{vec} = \prod_{(l,i)\neq(k,j)} \frac{f\big(h^{(l)}_i - h^{(k)}_j +(i-j)(1-\beta) + \beta \big)}{f \big(h^{(l)}_i - h^{(k)}_j +(i-j)(1-\beta) \big)} \frac{f \big(b_{lk} + \beta(j - i) \big)}{f\big(b_{lk} + \beta(j - i) + \beta \big)},     \label{Zff-app}
\ee
in a form suitable for matrix model interpretation. To start with, we restrict to two sets of partitions, $n=2$. In this case we denote the 2-tuple of partitions as
$$
(\lambda,\mu) \equiv (\lambda^{(1)},\lambda^{(2)}) = \vec{\lambda},
$$
and instead of general (\ref{hl}) we introduce 
\be
h_i \equiv h^{(1)}_i = \lambda_i - i + N + b_1,  \qquad \qquad k_i \equiv h^{(2)}_i = \mu_i - i + N + b_2.    \label{hk}
\ee
Then (\ref{Zff-app}) takes the form
\be
Z_{n=2} = Z_{hh} Z_{kk} Z_{hk} Z_{kh}.   \label{Zff-n2}
\ee
where the consecutive terms arise respectively from the products in (\ref{Zff-n2}) over $l=k=1$, $l=k=2$, $(l=1, k=2)$ and $(l=2,k=1)$, which we consider now one by one. 

Firstly we consider $Z_{hh}$, which reads
\be
Z_{hh} = \prod_{i \neq j} \frac{f\big(h_i - h_j +(i-j)(1-\beta) + \beta \big)}{f \big(h_i - h_j +(i-j)(1-\beta) \big)} \frac{f \big( \beta(j - i) \big)}{f\big(\beta(j - i) + \beta \big)} \equiv Z^{<}_{hh}\, Z^{>}_{hh}, 
\ee
and $Z^{<}_{hh}$ and $Z^{>}_{hh}$ correspond to products with, respectively, $i<j$ and $i>j$. We consider first $Z^{<}_{hh}$ and split the product into three parts: $1\leq i<j\leq N$, $(1\leq i \leq N, N< j)$, and $N<i<j$. This leads to
\bea
Z^{<}_{hh} & = &  \prod_{i < j}^N \frac{f\big(h_i - h_j +(i-j)(1-\beta) + \beta \big)}{f \big(h_i - h_j +(i-j)(1-\beta) \big)} \times \nonumber \\
& & \qquad \qquad \times \prod_{i=1}^N \prod_{i=1}^{\infty} \frac{f\big(h_i-b_1+(i-N)(1-\beta) + j\beta + \beta\big)}{f\big(h_i-b_1+(i-N)(1-\beta) + j\beta\big)}  \frac{f(\beta j)}{f(\beta j+\beta)} \Big) = \nonumber \\
& = & \Big( \prod_{i < j}^N \frac{f\big(h_i - h_j +(i-j)(1-\beta) + \beta \big)}{f \big(h_i - h_j +(i-j)(1-\beta) \big)} \Big) \Big(\prod_{i=1}^N \frac{f(\beta)}{f\big(h_i-b_1+(i-N)(1-\beta)+\beta\big)} \Big),    \nonumber
\eea 
where we used $\prod_{j=1}^{\infty} \frac{f(x+j\beta+\beta)}{f(x+j\beta)} = \frac{1}{f(x+\beta)}$. In the same way we get
\be
Z^{>}_{hh} = \Big( \prod_{i > j}^N \frac{f\big(h_i - h_j +(i-j)(1-\beta) + \beta \big)}{f \big(h_i - h_j +(i-j)(1-\beta) \big)} \Big) \Big(\prod_{i=1}^N \frac{f\big(-(h_i-b_1+(i-N)(1-\beta))\big)}{f(0)} \Big).
\ee

The factor $Z_{kk}$ can be written in the same way as above, with $h_i$ and $b_1$ replaced respectively by $k_i$ and $b_2$
$$
Z_{kk} \equiv Z^{<}_{kk}\, Z^{>}_{kk} = Z^{<}_{hh} \, Z^{>}_{hh} \Big\vert_{h_i\to k_i,b_1\to b_2}.
$$

Next we consider $Z_{hk}$, which we write as
\be
Z_{hk} = \prod_{i \neq j} \frac{f\big(h_i - k_j +(i-j)(1-\beta) + \beta \big)}{f \big(h_i - k_j +(i-j)(1-\beta) \big)} \frac{f \big( \beta(j - i) + b_{12} \big)}{f\big(\beta(j - i) + \beta + b_{12} \big)} \equiv Z^{=}_{hk} \, Z^{<}_{hk}\, Z^{>}_{hk}.
\ee
The three factors $Z^{=,<,>}_{hk}$ on the right correspond respectively to $i=j$, $i<j$ and $i>j$, and similar manipulations as above lead to the following results
\bea
Z^{=}_{hk} & = & \Big(\prod_{i=1}^{N} \frac{f\big(h_i - k_i + \beta \big)}{f \big(h_i - k_i \big)} \Big) \Big(\frac{f(b_{12})}{f(b_{12}+\beta)}\Big)^N, \nonumber\\
Z^{<}_{hk} & = & \Big( \prod_{i < j}^N \frac{f\big(h_i - k_j +(i-j)(1-\beta) + \beta \big)}{f \big(h_i - k_j +(i-j)(1-\beta) \big)} \Big) \Big(\prod_{i=1}^N \frac{f(b_{12}+\beta)}{f\big(h_i-b_2+(i-N)(1-\beta)+\beta\big)} \Big), \nonumber \\
Z^{>}_{hk} & = & \Big( \prod_{i > j}^N \frac{f\big(h_i - k_j +(i-j)(1-\beta) + \beta \big)}{f \big(h_i - k_j +(i-j)(1-\beta) \big)} \Big) \Big(\prod_{i=1}^N \frac{f\big(-(k_i-b_1+(i-N)(1-\beta))\big)}{f(b_{12})} \Big).   \nonumber
\eea
Note that in the overall product of these three factors, all terms which involve $f(b_{12})$ and $f(b_{12} + \beta)$ cancel.

The factor $Z_{kh}$ can be obtained from $Z_{hk}$ by the substitution $h_i\leftrightarrow k_i$ and $b_1 \leftrightarrow b_2$
$$
Z_{kh} \equiv Z^{=}_{kh} \, Z^{<}_{kh}\, Z^{>}_{kh} = Z^{=}_{hk} \, Z^{<}_{hk}\, Z^{>}_{hk}\Big\vert_{h_i\leftrightarrow k_i, b_1 \leftrightarrow b_2}.
$$

The expression (\ref{Zff-n2}) is a product of the above factors, and from their form it is clear that it easily generalizes to arbitrary $n$: one has just to consider the analogous factors as above for all possible pairs $(l,k)$, with $l,k=1,\ldots,n$. This expression has also a simpler representation in terms of
$$
l_{i=1,\ldots,nN} := (h^{(n)}_1,\ldots,h^{(n)}_N,\ldots \ldots,h^{(1)}_1,\ldots,h^{(1)}_N) \qquad \Leftrightarrow \qquad l_{(k-1)N+i}:=h^{(n-k+1)}_i.
$$
In this notation, and denoting $i_{mod\, N}\equiv i\, mod\, N$, the expression (\ref{Zff-app}) for arbitrary $n$ takes the form
\bea
Z_{vec} & = & \Big( \frac{f(\beta)}{f(0)} \Big)^{nN} \,  \prod_{i \neq j}^{nN} \frac{f\big(l_i - l_j +(i_{mod\, N}-j_{mod\, N})(1-\beta) + \beta \big)}{f \big(l_i - l_j +(i_{mod\, N}-j_{mod\, N})(1-\beta) \big)}   \times \nonumber \\
& & \qquad  \times \prod_{l=1}^n  \prod_{i=1}^{nN}  \frac{f\big(-(l_i - b_l + (i_{mod\, N}-N)(1-\beta)) \big)}{f\big(l_i - b_l + (i_{mod\, N}-N)(1-\beta) + \beta \big)}.  
 \label{Zff-l-app}
\eea


\section{Asymptotics}    \label{app-asymptote}

\subsection{Asymptotics of gamma function and quantum dilogarithm}

The following expansion of the logarithm of the gamma function holds \cite{AbramSteg}

\be
\log \Gamma(z) = -\frac{1}{2}\log |z| + z\log |z| - z + \frac{1}{2}\log 2\pi + \sum_{n=1}^{\infty} \frac{B_{2n}}{2n(2n-1)z^{2n-1}} .   \label{lnGamma}
\ee

We also consider the asymptotics of the quantum dilogarithm, in the notation of \cite{eynard-planch,SW-matrix} written as
\be
g(z) = \prod_{i=1}^{\infty} \big(1 - \frac{1}{z}q^i\big) = \big(\frac{q}{z};q\big)_{\infty},      \label{g-dilog}
\ee   
where the $q$-Pochhammer symbol is
$$
(z;q)_{\infty} = \prod_{i=0}^{\infty} (1-zq^i).
$$
For $q=e^{\epsilon_2}$ we have
\be
\log g(z) = \frac{1}{\epsilon_2} \sum_{m=0}^{\infty} \Li_{2-m}\big(\frac{1}{z}\big) \frac{B_m}{m!} (-\epsilon_2)^m.        \label{g-dilog-asymp}
\ee

\subsection{Asymptotics of the ratio of gamma functions}   \label{app-ratios}

The asymptotics
\be
\frac{\Gamma(z+\alpha+\beta)}{\Gamma(z+\alpha)} = z^{\beta} \Big(1 + \sum_{n=1}^{\infty} C_n(\beta,\alpha) z^{-n} \Big)   \label{asympt-Gamma}
\ee
was derived in \cite{GammaRatio}. The first term of this expansion \cite{AbramSteg} can be obtained from the Stirling formula, though to get the entire expansion requires more work. The coefficients $C_n$ are given by 
$$
C_n(\beta,\alpha) = \sum_{m=0}^n {\beta-m  \choose n-m} A_m(\beta) \alpha^{n-m},
$$
where coefficients $A_m$ satisfy the recursion formula
$$
A_n(\alpha) = \frac{1}{n} \sum_{m=0}^{n-1} {\alpha-m  \choose  n-m+1} A_m(\alpha),
$$
and we also set $C_0=1$. From this prescription one can find
$$
C_1(\beta,\alpha) = \frac{1}{2} \beta (\beta+2\alpha-1), \quad C_2(\beta,\alpha) = \frac{1}{12} {\beta \choose 2} \Big((\beta-2)(3\beta-1) + 12 \alpha(\beta+\alpha-1) \Big), \quad \textrm{etc.}
$$

One can also write 
$$
C_n(\beta,\alpha) = \frac{c_n}{\Gamma(\beta-n+1)},
$$
where $c_n$ are given in terms of the generating function
$$
\Gamma(1+\beta) e^{(\alpha+\beta)t} (e^t-1)^{-1-\beta} \equiv \sum_{n=0}^{\infty} c_n t^{n-1-\beta}.
$$


\subsection{Asymptotics of the ratio of quantum dilogarithms}

In the five-dimensional case we also need the following asymptotics
\be
\frac{(zq^{\alpha};q)_{\infty}}{(zq^{\alpha+\beta};q)_{\infty}} \simeq (1-z)^{\beta},  \label{asympt-qGamma}
\ee
which can be found explicitly e.g. in \cite{Gasper}. This formula is closely related to the ratio of $q$-gamma functions $\G_q(x)$. The function $\Gamma_q$ is defined as
$$
\Gamma_q(x) = (1-q)^{1-x}\frac{(q;q)_{\infty}}{(q^x;q)_{\infty}},
$$
and it is known that
$$
\lim_{q\to 1} \Gamma_q(x) = \Gamma(x).
$$
This function can also be expressed as the $q$-hypergeometric function. One can find the asymptotics (\ref{asympt-qGamma}) using the $q$-analogue of the Stirling formula for $q$-gamma function, which is derived in \cite{Moak} (see also \cite{Mansour}).


\section{Matrix model potentials}    \label{app-V4dvec}

As explained in the introduction, our methods allow to get the leading $\epsilon_2$ terms of matrix model potentials. These leading contributions should be sufficient to get the spectral curve of the matrix model, and subsequently solve it. The subleading terms, for general $\beta$, depend not only on $\lambda^{(l)}_i$, but also explicitly on $(i\,mod\, N)$. Therefore they cannot be simply symmetrized and their matrix model interpretation is less clear. Nonetheless, for completeness, we write down the full form of the four-dimensional potential (\ref{V4dvec-leading}) arising from $Z^{4d}_{\vec{\lambda},vec}$. Up to the symmetrization of eigenvalues it reads:
\bea
V^{4d}_{vec}(x) & = & \sum_{l=1}^n \Big[-2\big(x-a_l + \epsilon_2(i_{mod \, N}-N)(1-\beta) \big) +  \nonumber \\
& & + \big(x-a_l + \epsilon_2(i_{mod \, N}-N)(1-\beta)\big) \times \nonumber \\
& & \qquad \qquad \times \Big(\log\big( x-a_l + \epsilon_2(i_{mod \, N}-N)(1-\beta)  \big) + \nonumber \\
& & \qquad \qquad \qquad \qquad + \log\big( x-a_l + \epsilon_2(i_{mod \, N}-N)(1-\beta) +\epsilon_2\beta \big)  \Big) + \nonumber \\
& & + \frac{\epsilon_2}{2} \log\big( x-a_l + \epsilon_2(i_{mod \, N}-N)(1-\beta)  \big) + \nonumber \\
& & - \frac{\epsilon_2}{2} \log\big( x-a_l + \epsilon_2(i_{mod \, N}-N)(1-\beta) + \epsilon_2\beta \big) + \nonumber \\
& & -\epsilon_2\beta + \epsilon_2\beta \log\big( x-a_l + \epsilon_2(i_{mod \, N}-N)(1-\beta) + \epsilon_2\beta \big) + \nonumber \\
& & + \sum_{i=1}^{\infty} \frac{B_i \epsilon_2^{2i}}{2i(2i-1)} \Big(\frac{1}{\big(x-a_l + \epsilon_2(i_{mod \, N}-N)(1-\beta)\big)^{2i-1}} + \nonumber \\
& & \qquad \qquad + \frac{1}{\big(x-a_l + \epsilon_2(i_{mod \, N}-N)(1-\beta) + \epsilon_2\beta\big)^{2i-1}}  \Big)
\Big]     \label{V4dvec-full}
\eea
This form arises from asymptotics presented in the appendix \ref{app-asymptote}, and contributions for other factors are analogous.



\newpage

\begin{center}
\begin{Large} {\bf Acknowledgments}  \end{Large}
\end{center}

\medskip

I am grateful to Robbert Dijkgraaf, Hiroaki Kanno, Albrecht Klemm, Nikita Nekrasov, Niclas Wyllard, and especially Bertrand Eynard for discussions, correspondence, and remarks on the manuscript. I thank George Gasper, Mourad Ismail, Yaming Yu and Ruiming Zhang for correspondence on various asymptotics. I also thank IPhT-CEA Saclay for hospitality. This research  was supported by the DOE grant DE-FG03-92ER40701FG-02, the Foundation for Polish Science, and the European Commission under the Marie-Curie International Outgoing Fellowship Programme. The contents of this publication reflect only the views of the author and not the views of the funding agencies. 


\end{document}